\documentclass[onecolumn,secnumarabic,amssymb, amsmath, nofootinbib,tightenlines,
nobibnotes, aps, prl,epsfig]{revtex4}
\usepackage{graphicx}
\usepackage{dcolumn}
\usepackage{bm}
\begin{document}
\preprint{APS/123-QED}
\title{An analysis of the longitudinal structure function at next-to-leading order
approximation at small-$x$}

\author{G.R.Boroun}%
 \email{boroun@razi.ac.ir }
 \affiliation{Department of Physics, Razi University, Kermanshah
67149, Iran}%
\author{Yanbing Cai }
\altaffiliation{yanbingcai@mail.gufe.edu.cn}
\affiliation{Guizhou Key Laboratory in Physics and Related Areas,
Guizhou University of Finance and Economics, Guiyang 550025, China}
\date{\today}
\begin{abstract}
The longitudinal structure function is considered at the
next-to-leading order approximation using the expansion method, as
defined by M.B.Gay Ducati and P.B.Goncalves [Phys.Lett.B {\bf390},
401 (1997)] and further developed by Jingxuan Chen et al.,
[Chin.Phys.C {\bf48}, 063104 (2024)]. This method provides results
for  a wide range of $x$ and $Q^2$ values. It is observed  that
the behavior of the longitudinal structure function  depends on
the fractional momentum carried by gluons at low $x$. The
extracted longitudinal structure functions $F_{L}(x,Q^2)$ are in
line with data from the H1 Collaboration [V. Andreev et al. (H1
Collaboration),
Eur. Phys. J. C 74, 2814 (2014)] and the CT18 parametrization method [T.-J. Hou et al., Phys. Rev. D 103, 014013 (2021)].\\

\end{abstract}
 \pacs{***}
\maketitle
\subsection{Introduction}

Quantum chromodynamics (QCD) is a theory describing the strong
interaction, in which the fundamental degrees of freedom are
quarks and gluons \cite{int1}. Determining the inner quark and
gluon constituents of the proton is crucial for QCD applications,
as it is a key ingredient in QCD factorization calculations.
Unlike the naive quark parton model \cite{int2}, which suggestes
that the proton is composed solely of quarks, QCD reveals a more
intricate structure for the proton, including not only quarks but
also anti-quarks and gluons. The probability distribution of these
constituents within the nucleon is known as parton distribution
functions (PDFs), defined as the probability density for finding a
parton with a longitudinal momentum fraction $x$ at resolution
scale $Q^2$. However, due to the non-perturbative nature of
partons, PDFs can not be calculated using perturbative QCD.
Fortunately, high-energy lepton-nucleon scattering offers a unique
opportunity to determine and test the partonic structure of the
proton \cite{int3}. Pioneering deep inelastic electron-proton
scattering (DIS) experiments conducted at SLAC have confirmed the
partonic substructure of the proton \cite{int4,int5}.
Additionally, DIS measurements by the H1 and ZEUS Collaborations
at HERA have been used to extract the proton$^{,}$s parton
distributions based on the factorization theorem \cite{int6}. The
future Electron-Ion Colliders (EIC) \cite{int7} and Electron-Ion
Colliders in China (EicC) \cite{int8} are expected to generate
more high-precision DIS data, significantly improving our
understanding of proton structure.

In the DIS process, the reduced neutral current differential cross
section can be expressed in terms of three proton structure
functions $F_2$, $xF_3$, and $F_L$ \cite{int3}. These structure
functions are related to the proton$^,$s parton distributions. The
electromagnetic structure function $F_2$ is associated with pure
photon exchange and represents the dominant contribution to the
cross section across most of the kinematic range. The structure
function $xF_3$ captures the difference between electron-proton
and positron-proton cross sections. Within the framework of the
quark parton model, $xF_3$ is directly related to valence-quark
distributions. Consequently, measuring $xF_3$ provides insights
into the lower-$x$ behavior of these valence-quark distributions.
The longitudinal structure function $F_L$ is proportional to the
cross section for interactions involving a longitudinally
polarized virtual photon interacting with a proton. Notably, at
leading order in QCD, $F_L$ vanishes; however, it may become
non-zero when gluon contributions are taken into account
\cite{int9}. In the small-$x$ region, the contribution from gluons
significantly surpasses that of quarks. This results in $F_L$
being directly sensitive to the gluon density, thereby serving as
a direct measure of the gluon distribution \cite{int10}.

The longitudinal structure function has been measured by the H1
and ZEUS collaborations \cite{int11,int12,int13,int14} and has
been updated in Ref. \cite{int15}, enhancing experimental
precision and extending coverage across a broader kinematic range.
These measurements provide significant constraints on the gluon
distribution within the proton, as the longitudinal structure
function can be obtained from the gluon distribution \cite{int16}.
In the literature, the longitudinal structure function has been
analytically examined through the gluon distribution which has
been expanded at \( z=0 \) \cite{int17} and at \( z=\alpha \)
(where \( 0 \leq \alpha < 1 \)) \cite{int18}. In our previous
paper, we derived an analytical gluon distribution at low $x$
based on the DGLAP equation, utilizing approximated leading-order
(LO) and next-to-leading-order (NLO) splitting functions
\cite{int19}. The analytical distribution provides a good
description of the differential structure function. In this paper,
we extend this analytical distribution to explore the longitudinal
structure function. We find that our analytical results are
consistent with data from the H1 Collaboration and the CT18
parametrization method.\\

\subsection{Method}

Recently, authors in Ref.\cite{int19} presented an analytical
solution for the derivative of the proton structure function with
respect to ${\ln}Q^2$ , denoted as
$\frac{{\partial}F_{2}(x,Q^2)}{{\partial}{\ln}Q^2}$. This solution
was derived using the gluon distribution at the leading-order (LO)
and the next-to-leading order(NLO) approximations in the small-$x$
limit. The analysis is based on the expansion method at the
expansion point $z=\alpha$ as reported in Refs.\cite{Ref2, Ref3,
Ref4}. In Ref.\cite{int19}, the gluon distribution from DGLAP
evolution equations \cite{Ref5, Ref6,Ref7} using the Mellin
transform was obtained. By neglecting the quark distribution at
small-$x$, the evolution of the gluon density at the NLO
approximation is defined by the following form
\begin{eqnarray}
\frac{{\partial}g(x,Q^{2})}{{\partial}{\ln}Q^{2}}=\frac{\alpha_{s}(Q^2)}{2\pi}
\int_{x}^{1}\bigg{[}2N+\frac{\alpha_{s}(Q^2)}{2\pi}(\frac{4}{3}C_{F}T_{f}
-\frac{46}{9}NT_{f}) \bigg{]}g(\frac{x}{z},Q^2)\frac{dz}{z^{2}},
\end{eqnarray}
where the splitting functions at the LO and NLO approximations, at
small fraction momentum, can be approximated as
\begin{eqnarray}
P_{gg}^{\mathrm{LO}+\mathrm{NLO}}|_{z{\ll}1}{\approx}\frac{1}{z}\bigg{[}2N+\frac{\alpha_{s}(Q^2)}{2\pi}(\frac{4}{3}C_{F}T_{f}
-\frac{46}{9}NT_{f})\bigg{]}.
\end{eqnarray}
For the SU(N) gauge group, we have $C_{A}=N$,
$C_{F}=\frac{N^2-1}{2N}$, $T_{f}=n_{f}T_{R}$ and $T_{R}=1/2$ where
$C_{F}$ and $C_{A}$ are the color Cassimir operators.\\
The DGLAP evolution equation for the gluon distribution function
can be rewritten into the Mellin transform using the fact that the
Mellin transform of a convolution factors is simply the ordinary
product of the Mellin transform of the factors by the following
form
\begin{eqnarray}
G_{\omega}(x,Q^2)=\exp(\frac{1}{\omega}\eta(Q_{0}^{2},
Q^2))G_{\omega}(x,Q_{0}^2),
\end{eqnarray}
where
\begin{eqnarray}
\eta(Q_{0}^{2},
Q^2))=\int_{Q_{0}^{2}}^{Q^2}\frac{dq^{2}}{q^{2}}\frac{\alpha_{s}(q^2)}{2\pi}\bigg{[}2N+\frac{\alpha_{s}(q^2)}{2\pi}(\frac{4}{3}C_{F}T_{f}
-\frac{46}{9}NT_{f})\bigg{]}.
\end{eqnarray}
The inverse Mellin transform of the coefficients is
straightforward as
\begin{eqnarray}
G(x,Q^2)=\int_{a-i\infty}^{a+i\infty}\frac{d\omega}{2{\pi}i}\exp(P(\omega))G_{\omega}(x,Q_{0}^2),
\end{eqnarray}
where
\begin{eqnarray}
P(\omega)=\omega{\ln}\frac{1}{x}+\frac{1}{\omega}\eta(Q_{0}^{2},
Q^2).
\end{eqnarray}
The analytical solution for the gluon distribution, after some
rearranging, is obtained in Ref.\cite{int19} in the following form
\begin{eqnarray}
G(x,Q^2)=K(Q^2)I_{0}(2(\rho{\ln}(1/x))^d),
\end{eqnarray}
where
\begin{eqnarray}
 K(Q^2)=a[\exp(\xi-\xi_{0})+b]\exp[c(\xi-\xi_{0})^{1/2}],
 \end{eqnarray}
with $\xi={\ln}\ln(Q^2/\Lambda^{2})$ and
$\xi_{0}={\ln}\ln(Q_{0}^2/\Lambda^{2})$ where $\Lambda$ is the QCD
cut-off parameter. The function $\rho$ at the LO approximation is
defined as
\begin{eqnarray}
\rho(Q^2)=\frac{4N}{\beta_{0}}{\ln}\frac{t}{t_{0}},
\end{eqnarray}
where $t={\ln}\frac{Q^2}{\Lambda^{2}}$,
$t_{0}={\ln}\frac{Q_{0}^2}{\Lambda^{2}}$ and
$\beta_{0}=\frac{1}{3}(33-2n_{f})$. At the NLO approximation the
function $\rho$ is modified by the following form
\begin{eqnarray}
\rho(Q^2)=\frac{4N}{\beta_{0}}{\ln}\frac{t}{t_{0}}R(t),
\end{eqnarray}
where
\begin{eqnarray}
R(t)=\frac{\widetilde{a}t^{\widetilde{d}}}{\widetilde{b}+\widetilde{c}t^{\widetilde{e}}}.
\end{eqnarray}
The coefficients at the LO and NLO approximations were determined
to fit the CJ15LO \cite{Ref8} and CJ15NLO \cite{Ref9} data with
$Q_{0}^2=1~\mathrm{GeV}^2$ and $2.5~\mathrm{GeV}^2$
respectively.\\

At small values of $x$, the longitudinal structure function is
driven mainly by gluons as the Altarelli and Martinelli
\cite{int16} equation for the gluonic longitudinal structure
function $F^{g}_{L}(x,Q^{2})$ is defined
\begin{eqnarray}
F^{g}_{L}(x,Q^{2})=<e^{2}>C_{L,g}(\alpha_{s}(Q^{2}),x){\otimes}xg(x,Q^{2})=<e^{2}>\int_{x}^{1}\frac{dz}{z}C_{L,g}(\alpha_{s}(Q^{2}),z)G(\frac{x}{z},Q^{2}),
\end{eqnarray}
where  $<e^{2}>$ is the average charge for the active quark
flavors, $<e^{2}>=n_{f}^{-1}\sum_{i=1}^{n_{f}}e_{i}^{2}$, and
$C_{L,g}$is the coefficient function that can be written by the
perturbative expansion at the LO and NLO approximations as follows
\cite{Ref11}
\begin{eqnarray}
 C_{L,g}(\alpha_{s},x)=\frac{\alpha_{s}}{4\pi}c_{L,g}^{\mathrm{LO}}(x)+(\frac{\alpha_{s}}{4\pi})^{2}c_{L,g}^{\mathrm{NLO}}(x),
\end{eqnarray}
where $c_{L,g}^{\mathrm{LO}}(z)=8n_{f}z^2(1-z)$ and
$c_{L,g}^{\mathrm{NLO}}(z)|_{z{\rightarrow}0}{\approx}{-5.333}n_{f}$.\\
Cooper-Sarkar et al in Ref.\cite{int17} have suggested that the
longitudinal structure function by expansion of the gluon density
around $z=0$ has the following form
\begin{eqnarray}
F_{L}(x,Q^2){\simeq}\frac{\alpha_{s}(Q^2)\sum_{f}e_{i}^{2}}{3\pi}\frac{6}{5.9}G(2.5x,Q^2).
\end{eqnarray}
Authors in Ref.\cite{int18} extended Eq.(12) based on expanding
the gluon distribution around $z=\alpha$ as the gluon distribution
can be expanded using the expansion method at an arbitrary point
$z=\alpha$ by the following form
\begin{eqnarray}
G(\frac{x}{1-z})|_{z=\alpha}&=&G(\frac{x}{1-\alpha})+\frac{x}{1-\alpha}(z-a)\frac{{\partial}G(\frac{x}{1-\alpha})}{{\partial}x}+\mathcal{O}(z-\alpha)^{2},
\end{eqnarray}
where the series is convergent for $|z-\alpha|<1$ as
 \begin{equation}
\frac{x}{1-z}|_{z=\alpha}=\frac{x}{1-\alpha}\sum_{k=1}^{\infty}[1+\frac{(z-\alpha)^{k}}{(1-\alpha)^{k}}].
\end{equation}
The longitudinal structure function at the LO approximation is
defined \cite{int18} by the following form
\begin{eqnarray}
F_{L}(x,Q^2){\simeq}\frac{\alpha_{s}(Q^2)\sum_{f}e_{i}^{2}}{3\pi}\frac{6}{5.9}G\bigg{(}\frac{x}{1-\alpha}\bigg{(}\frac{3}{2}-\alpha\bigg{)},Q^2\bigg{)},
\end{eqnarray}
where this result is similar to Eq.(14) when the expansion point
$\alpha=0.666$ is used.\\
In the NLO approximation, the longitudinal structure function is
given by
\begin{eqnarray}
F_{L}(x,Q^2){\simeq}F^{\mathrm{LO}}_{L}(x,Q^2)(\mathrm{i.e.},
\mathrm{Eq}.(10)){-5.333}\sum_{i=1}^{n_{f}}e_{i}^{2}\bigg{(}\frac{\alpha_{s}}{4\pi}\bigg{)}^{2}\int_{0}^{1-x}\frac{dz}{1-z}G(\frac{x}{1-z},Q^2).
\end{eqnarray}
The integral in Eq.(18) is modified by the following form
\begin{eqnarray}
\int_{0}^{1-x}\frac{dz}{1-z}G(\frac{x}{1-z},Q^2)=\frac{1}{x}\int_{0}^{1-x}{d\widetilde{z}}\widetilde{G}(\frac{x}{1-\widetilde{z}},Q^2),
\end{eqnarray}
where $\widetilde{G}(x,Q^2)=xG(x,Q^2)$. In the limit
$x{\rightarrow}0$, the expansion of $\widetilde{G}$ around the
point $z=\alpha$ gives \cite{int19}
\begin{eqnarray}
\frac{1}{x}\int_{0}^{1-x}{d\widetilde{z}}\widetilde{G}(\frac{x}{1-\widetilde{z}},Q^2)=
(1-x)\frac{\frac{3}{2}-2\alpha}{(1-\alpha)^2}G\bigg{(}\frac{\frac{3}{2}-2\alpha}{(1-\alpha)^2}x,Q^2
\bigg{)}.
\end{eqnarray}
Therefore, the gluonic longitudinal structure function in the  NLO
approximation is find
\begin{eqnarray}
F_{L}(x,Q^2){\simeq}\frac{\alpha_{s}(Q^2)\sum_{f}e_{i}^{2}}{3\pi}\frac{6}{5.9}G\bigg{(}\frac{x}{1-\alpha}\bigg{(}\frac{3}{2}-\alpha\bigg{)},Q^2\bigg{)}
{-5.333}\sum_{i=1}^{n_{f}}e_{i}^{2}\bigg{(}\frac{\alpha_{s}}{4\pi}\bigg{)}^{2}(1-x)\frac{\frac{3}{2}
-2\alpha}{(1-\alpha)^2}G\bigg{(}\frac{\frac{3}{2}-2\alpha}{(1-\alpha)^2}x,Q^2\bigg{)}.
\end{eqnarray}
We observe that the longitudinal structure function is dependent
on the gluon distribution at the expansion point $z=\alpha$.
Indeed, the free parameter $\alpha$ relates the gluon density to
the observable longitudinal structure function. Comparing with the
high precision data for the longitudinal structure function at
small $x$ help to confirm the effectiveness of the approach and
clarify what is the best value for the expansion point $\alpha$ in
the next section.\\

\subsection{Results and Discussion}

The running coupling at the LO and NLO approximations are defined
by the following forms
\begin{eqnarray}
\alpha_{s}(Q^2)&=&\frac{4\pi}{\beta_{0}{\ln}(Q^2/\Lambda^2)}\hspace{3.5cm}(\mathrm{LO})\nonumber\\
\alpha_{s}(Q^2)&=&\frac{4\pi}{\beta_{0}{\ln}(Q^2/\Lambda^2)}\bigg{[}
1-\frac{\beta_{1}{\ln}{\ln}(Q^2/\Lambda^2)}{\beta^{2}_{0}{\ln}(Q^2/\Lambda^2)}\bigg{]}~(\mathrm{NLO}),
\end{eqnarray}
with $\beta_{0}$ and $\beta_{1}$ as the first two coefficients of
the QCD $\beta$-function
\begin{eqnarray}
\beta_{0}&=&\frac{1}{3}(11C_{A}-2n_{f})\nonumber\\
\beta_{1}&=&\frac{1}{3}(34C^{2}_{A}-2n_{f}(5C_{A}+3C_{F})),
\end{eqnarray}
where $C_{F}=\frac{N_{c}^{2}-1}{2N_{c}}$ and $C_{A}=N_{c}$ are the
Casimir operators in the fundamental and adjoint representations
of the $\mathrm{SU(N_{c})}$ color group with $N_{c}=3$. $\Lambda$
is the QCD cut-off parameter and has been extracted from ZEUS data
with $\alpha_{s}(M_{Z}^{2})=0.1166$. The following results have
been obtained for $\Lambda$ with four active flavor numbers (i.e.,
$n_{f}=4$ ) as \cite{Ref14}
\begin{eqnarray}
\Lambda^{\mathrm{LO}}=136.8~\mathrm{MeV},~~~~~~\Lambda^{\mathrm{NLO}}=284.0~\mathrm{MeV}.
\end{eqnarray}
The longitudinal structure function $F_{L}(x,Q^2)$ with the
explicit form of the gluon density at the LO and NLO
approximations is extracted due to the expansion points
$0{\leq}~\alpha<1$. To present more detailed discussions on our
findings, the results for the longitudinal structure function are
compared with the CT18 \cite{Ref15} parametrization model in the
general mass-variable flavor number scheme (GM-VFNS) and in  the
zero mass-variable flavor number scheme (ZM-VFNS). The GM-VFNS is
dependent on the rescaling variable $\chi$\footnote{In order to
account for the production threshold effect for the charm quark at
$n_{f}=4$, one should consider the quark mass for small $Q^2$.}
where $\chi=x\bigg{(}1+\frac{4m_{c}^{2}}{Q^2}\bigg{)}$, as the
rescaling variable reduces to the Bjorken variable $x$ at high
$Q^2$ values. The data are taken from the H1-Collaboration
\cite{int15} at HERA in the region
$1.5{\leq}Q^2{\leq}800~\mathrm{GeV}^2$ with a lepton beam energy
of $27.6~\mathrm{GeV}$ and two proton beam energies of
$E_{p}=460~\mathrm{and}~575~ \mathrm{GeV}$ corresponding to
center-of-mass (COM) energies of $225$ and $252~ \mathrm{GeV}$,
respectively from the inclusive ep double differential cross
sections for neutral current deep inelastic scattering.\\
We have calculated the $x$-dependence of the longitudinal
structure function at several fixed values of $Q^2$ corresponding
to H1-Collaboration data in a wide range of the expansion point
$z=\alpha$. Results are presented in Figure 1 where the
$x$-evolution of $F_{L}(x,Q^2)$ is clearly exhibited for
$\alpha=0.0$ (red circles), $\alpha=0.5$ ( brown down-triangles)
and $\alpha=0.9$ ( green up-triangles). It is seen that, for low
values of the presented $Q^2$, the extracted longitudinal
structure function  within the NLO approximation is in much better
agreement with the CT18 parametrization model at $\alpha<0.5$.
Also, for high values of the presented $Q^2$, the extracted
longitudinal structure function within the NLO approximation is in
a much better agreement with the CT18 parametrization model at
$\alpha>0.5$. We have found that the choice of $\alpha\sim 0.0$
and $\alpha\sim 0.9$ for the expansion points at low and high
$Q^2$ values gives results that are sufficiently
accurate for our purposes respectively.\\
\begin{figure}
\centerline{
\includegraphics[width=0.8\textwidth]{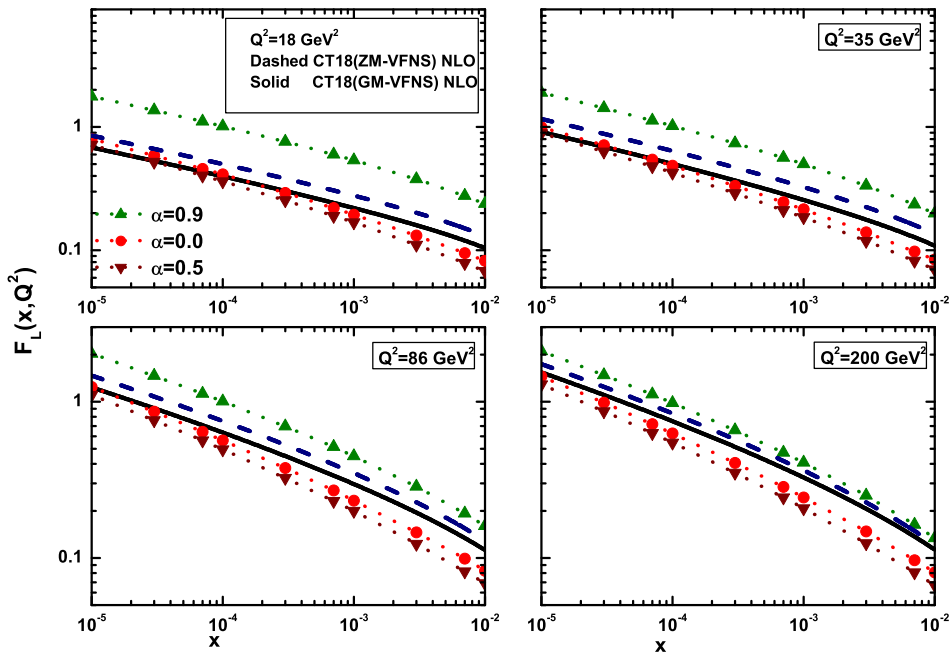}}
\caption{The longitudinal structure function $F_{L}(x,Q^2)$
plotted at fixed $Q^2$ as a function of $x$ variable for
$\alpha=0.0$ (red circles), $\alpha=0.5$ ( brown down-triangles)
and $\alpha=0.9$ ( green up-triangles), compared with the CT18
parametrization method \cite{Ref15} in the GM-VFNS (solid curves)
and in the ZM-VFNS (dashed curves) at the NLO
approximation.}\label{Fig1}
\end{figure}
In Fig.2, we present the $x$ dependence of the longitudinal
structure function at $Q^2=8.5$ and $200~\mathrm{GeV}^2$ compared
with the H1 Collaboration data \cite{int15} accompanied by total
errors and the results from CT18 (GM-VFNS) NLO parametrization
model. We observe that, with respect to the expansion points used
in the gluon density, the extracted longitudinal structure
functions within the NLO approximation are comparable with the
experimental data and the CT18 NLO model.\\
\begin{figure}
\centerline{
\includegraphics[width=0.6\textwidth]{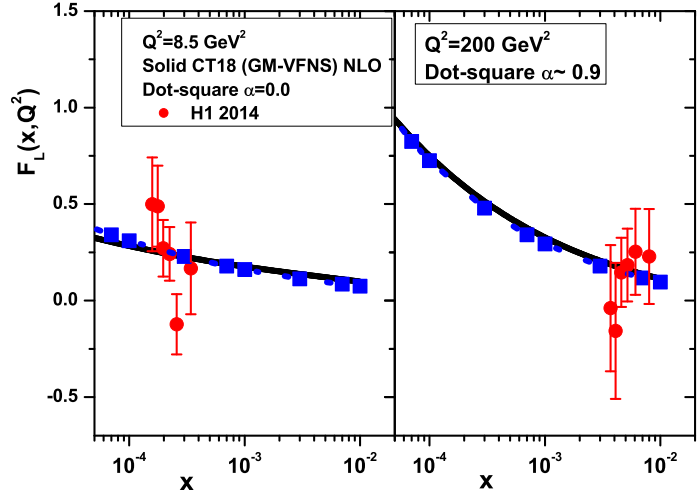}}
\caption{The longitudinal structure functions at the NLO
approximation with respect to the expansion points extracted in
comparison with the H1 experimental data (red circules)
\cite{int15} accompanied with total errors and the CT18 NLO
\cite{Ref15} parametrization model.}\label{Fig2}
\end{figure}
In Fig. 3, we show the $Q^2$ dependence of the longitudinal
structure function at small $x$ at the NLO approximation at fixed
values of the invariant mass $W$ as $W=230~\mathrm{GeV}$. In this
figure (i.e., Fig. 3), the results of calculations at the
expansion points $\alpha=0.0$ and $0.9$ and the comparison with
the H1 Collaboration data \cite{int15} accompanied by total errors
are presented. The extracted values in a wide range of the
expansion points are in good agreement with experimental data at
low and high $Q^2$ values.\\
\begin{figure}
\centerline{
\includegraphics[width=0.6\textwidth]{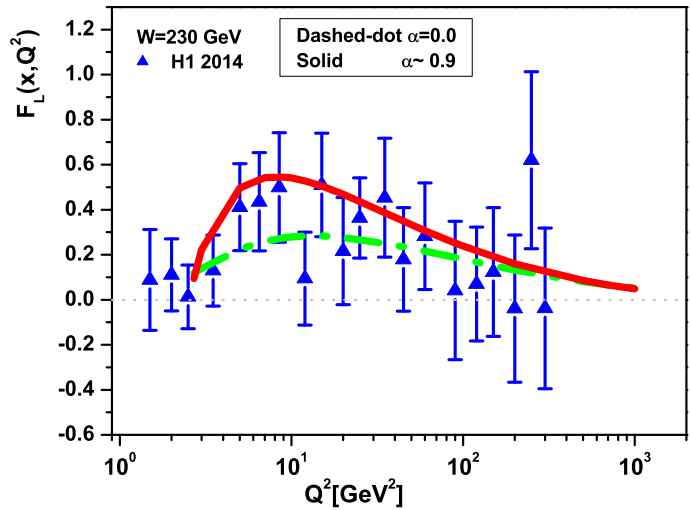}}
\caption{The extracted longitudinal structure functions
$F_{L}(x,Q^2)$  at a fixed value of the invariant mass $W$ as
$W=230~\mathrm{GeV}$ at the expansion points $\alpha=0.0$
(dashed-dot curve) and $0.9$ (solid curve) compared with the H1
Collaboration data \cite{int15} accompanied with total
errors.}\label{Fig3}
\end{figure}

In conclusion, we have presented a method based on the expansion
of gluon density to determine the longitudinal structure function
at the NLO approximation at low $x$. This method relies on the
momentum carried by the gluon density due to the expansion points
within a kinematical region characterized by low values of the
Bjorken variable $x$. We find that the expansion method of gluon
density provides correct behaviors of the extracted longitudinal
structure function $F_{L}(x,Q^2)$ at low and high $Q^2$ values due
to the expansion points $\alpha\simeq 0.0$ and $0.9$ and that our
results for $F_{L}(x,Q^2)$ demonstrate comparability with data
from the H1 Collaboration, the CT18 parametrization method and
other results obtained using the momentum space method \cite{Ref17}.\\
\subsection{ACKNOWLEDGMENTS}
G.R.Boroun is grateful to Razi University for the financial
support of this project.




\end{document}